\documentclass{revtex4}
\usepackage{amssymb}
\usepackage{amsmath}
\usepackage{graphicx}
\usepackage[font={footnotesize,it}]{caption}

\input{tcilatex}

\begin{document}

\title{Rindler type acceleration in f(R) gravity }
\author{S. Habib Mazharimousavi}
\email{habib.mazhari@emu.edu.tr}
\author{M. Halilsoy}
\email{mustafa.halilsoy@emu.edu.tr}
\affiliation{Physics Department, Eastern Mediterranean University, G. Magusa north
Cyprus, Mersin 10 Turkey}

\begin{abstract}
By choosing a fluid source in $f(R)$ gravity, defined by $f\left( R\right)
=R-12a\xi \ln \left\vert R\right\vert $, where $a$ (=Rindler acceleration)
and $\xi $ are both constants, the field equations correctly yield the
Rindler acceleration term in the metric. We identify domains in which the
weak energy conditions (WEC) and the strong energy conditions (SEC) are
satisfied.
\end{abstract}

\maketitle

Rindler acceleration is known to act on an observer accelerated in flat
spacetime. Geometrically such a spacetime is represented by $%
ds^{2}=-x^{2}dt^{2}+dx^{2}+dy^{2}+dz^{2},$ where the acceleration in
question acts in the $x-$direction \cite{1}. The reason that this
acceleration has become popular in recent years anew is due to an analogous
effect detected in the Pioneer spacecrafts launched in 1972 / 73.
Observation of the spacecrafts over a long period revealed an attractive,
mysterious acceleration toward Sun, an effect came to be known as the
Pioneer anomaly \cite{2}. Besides the MOdified Newton Dynamics (MOND) \cite%
{3} to account for such an extraneous acceleration there has been attempts
within general relativity for a satisfactory interpretation. From this token
a field theoretical approach based on dilatonic source in general relativity
was proposed by Grumiller to yield a Rindler type acceleration in the
spacetime \cite{4,5}. More recently we attempted to interpret the Rindler
acceleration term as a non-linear electrodynamic effect with an unusual
Lagrangian \cite{6}. Therein the problematic energy conditions are satisfied
but at the cost of extra structures such as global monopoles \cite{7} which
pop up naturally. In a different study global monopoles were proposed as
source to create the acceleration term in the weak field approximation \cite%
{8}.

In this Letter we show that a particular $f(R)$ gravity \cite%
{9,10,11,12,13,14} with a fluid source accounts for the Rindler
acceleration. The fluid satisfies the weak energy condition (WEC) and strong
energy condition (SEC) in regions as depicted in Fig. 1.

The action for $f(R)$ gravity written as%
\begin{equation}
S=\frac{1}{2\kappa }\int \sqrt{-g}f\left( R\right) d^{4}x+S_{M}
\end{equation}%
in which $\kappa =8\pi G=1,$ $f\left( R\right) =R-12a\xi \ln \left\vert
R\right\vert ,$ ($a$ and $\xi $ are constants) is a function of the Ricci
scalar $R$ and $S_{M}$ is the physical source for a perfect fluid-type
energy momentum 
\begin{equation}
T_{\mu }^{\nu }=diag.\left[ -\rho ,p,q,q\right]
\end{equation}%
with a state function $p=-\rho .$ Note that for dimensional reasons the
logarithmic argument should read $\left\vert \frac{R}{R_{0}}\right\vert $
where $\ln \left\vert R_{0}\right\vert $ accounts for the cosmological
constant. In our analysis, however, we shall choose $\left\vert
R_{0}\right\vert =1$ to ignore the cosmological constant. The $4-$%
dimensional static spherically symmetric line element is given by%
\begin{equation}
ds^{2}=-A\left( r\right) dt^{2}+\frac{1}{B(r)}dr^{2}+r^{2}\left( d\theta
^{2}+\sin ^{2}\theta d\varphi ^{2}\right)
\end{equation}%
where $A\left( r\right) $ and $B\left( r\right) $ are to be found. Let us
add also that in the sequel, for convenience we shall make the choice $%
A\left( r\right) =B(r)$.

Variation of the action with respect to the metric yields the field equations%
\begin{equation}
G_{\mu }^{\nu }=\frac{1}{F}T_{\mu }^{\nu }+\check{T}_{\mu }^{\nu }
\end{equation}%
where $G_{\mu }^{\nu }$ stands for the Einstein's tensor, with~%
\begin{equation}
\check{T}_{\mu }^{\nu }=\frac{1}{F}\left[ \nabla ^{\nu }\nabla _{\mu
}F-\left( \square F-\frac{1}{2}f+\frac{1}{2}RF\right) \delta _{\mu }^{\nu }%
\right] .
\end{equation}%
Our notation here is such that $\square =\nabla ^{\mu }\nabla _{\mu }=\frac{1%
}{\sqrt{-g}}\partial _{\mu }\left( \sqrt{-g}\partial ^{\mu }\right) $ and $%
\nabla ^{\nu }\nabla _{\mu }h=g^{\lambda \nu }\nabla _{\lambda }h_{,\mu
}=g^{\lambda \nu }\left( \partial _{\lambda }h_{,\mu }-\Gamma _{\lambda \mu
}^{\beta }h_{,\beta }\right) $ for a scalar function $h$. The field
equations explicitly read as%
\begin{equation}
FR_{t}^{t}-\frac{f}{2}+\square F=\nabla ^{t}\nabla _{t}F+T_{t}^{t}
\end{equation}%
\begin{equation}
FR_{r}^{r}-\frac{f}{2}+\square F=\nabla ^{r}\nabla _{r}F+T_{r}^{r}
\end{equation}%
\begin{eqnarray}
FR_{\theta }^{\theta }-\frac{f}{2}+\square F &=&\nabla ^{\theta }\nabla
_{\theta }F+T_{\theta }^{\theta } \\
&&\left( F=\frac{df}{dR}\right) 
\end{eqnarray}%
which are independent. Note that the $\varphi \varphi $ equation is
identical with $\theta \theta $ equation. By adding the four equations
(i.e., $tt$, $rr$, $\theta \theta $ and $\varphi \varphi $) we find 
\begin{equation}
FR-2f+3\square F=T
\end{equation}%
which is the trace of Eq. (4). As usual $tt$ and $rr$ components admit $%
\nabla ^{t}\nabla _{t}F=\nabla ^{r}\nabla _{r}F$ which in turn yields $%
F^{\prime \prime }=0,$ with prime $^{\prime }=\frac{d}{dr},$ and
consequently 
\begin{equation}
F=C_{1}+C_{2}r.
\end{equation}%
Here $C_{1}$ and $C_{2}$ are two integration constants which for our purpose
we set $C_{1}=1$ and $C_{2}=\xi .$ A detailed calculation gives the metric
solution 
\begin{equation}
A\left( r\right) =1-\frac{2m}{r}+2ar
\end{equation}%
which would lead to the following energy-momentum components%
\begin{equation}
-\rho =p=\frac{\left( 6a\xi -f\right) r^{2}+4\left( \xi -a\right) r-6m\xi }{%
2r^{2}},
\end{equation}%
\begin{equation}
q=-\frac{fr-2\xi +8a}{2r}
\end{equation}%
with 
\begin{equation}
f=f\left( R\left( r\right) \right) =-\left( \frac{12a}{r}+12a\xi \ln \left( 
\frac{12a}{r}\right) \right) .
\end{equation}%
It is observed that in the limit of $R-$gravity (i.e., $\xi \rightarrow 0$)
one gets $f=R$ $=-\frac{12a}{r}$ and therefore 
\begin{equation}
-\rho =p=\frac{4a}{r}
\end{equation}%
while 
\begin{equation}
q=\frac{2a}{r}.
\end{equation}%
Naturally the integration constant $m$ accounts for the constant mass while
the Rindler acceleration constant, i.e. $a$, determines the properties of
the fluid source. These are the results found in \cite{4,5}. In the other
limit, once $a\rightarrow 0$ one can see from the vanishing Ricci scalar
that $F$ can not be $r$ dependent which means that $\xi $ must be zero.
This, in turn, reduces to the standard $R-$gravity.

Once more we note that the Rindler acceleration $a$ is positive and $%
C_{2}=\xi $ is positive too, to avoid any non-physical solutions. Our final
remark will be on the energy conditions.

For WEC one should have (i) $\rho \geq 0,$ (ii) $\rho +p\geq 0$ and (iii) $%
\rho +q\geq 0.$ One observes that the second condition is trivial while the
third condition implies%
\begin{equation}
3a\xi r^{2}+\left( 2a+\xi \right) r-3m\xi \leq 0
\end{equation}%
which simply confines the range of $r$ as 
\begin{equation}
r\leq \frac{\sqrt{\left( 2a+\xi \right) ^{2}+36ma\xi ^{2}}-\left( 2a+\xi
\right) }{6a\xi }.
\end{equation}%
On the other hand the first condition ($\rho \geq 0$) reads as%
\begin{equation}
-6a\xi r^{2}\ln \left( \frac{12a}{r}\right) -\left( \xi +2a\right) r\leq
3a\xi r^{2}+\left( 2a+\xi \right) r-3m\xi \leq 0
\end{equation}

\begin{figure}[tbp]
\includegraphics[width=130mm,scale=1]{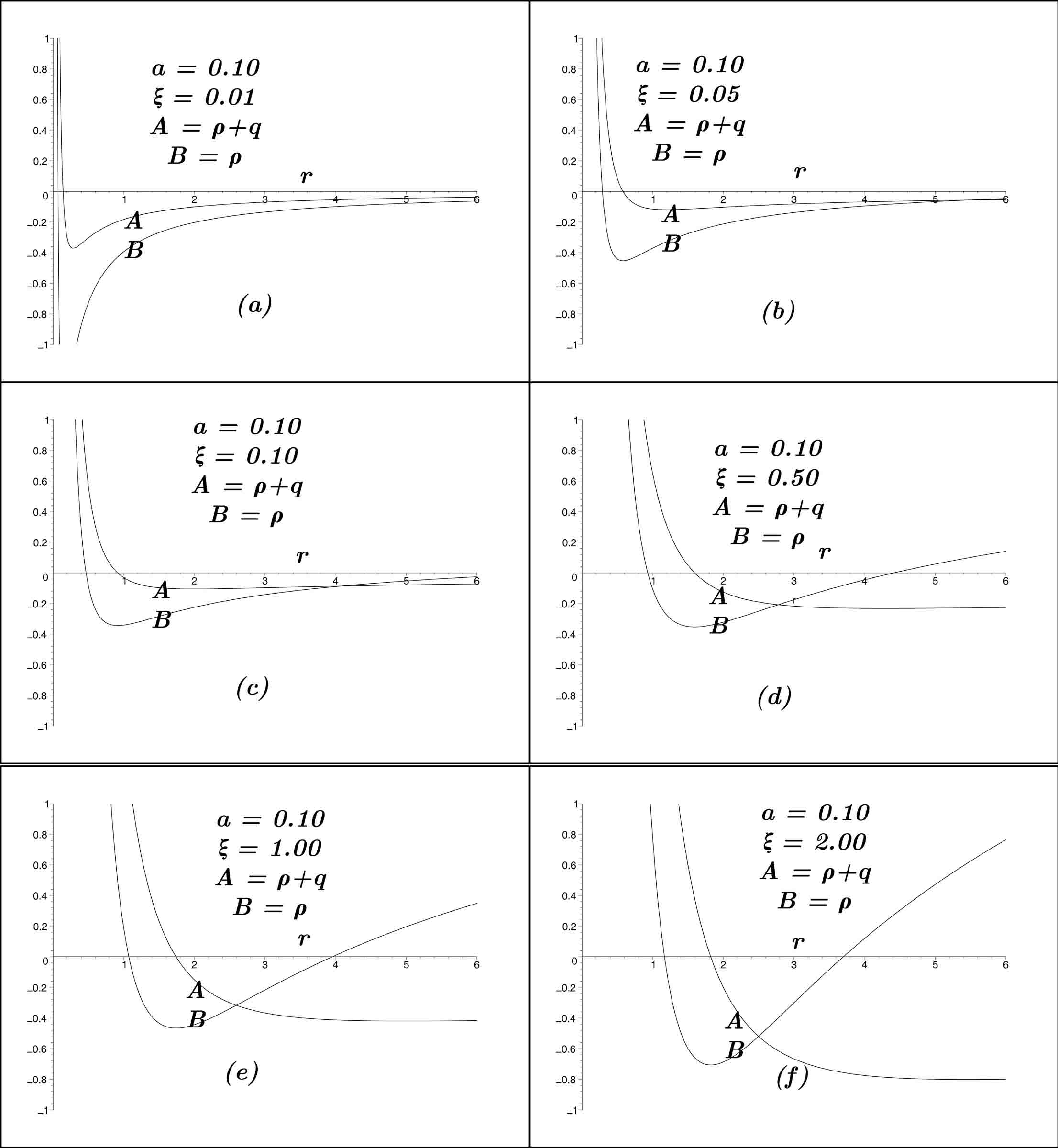}
\caption{A plot of $\protect\rho $ (=B) and $\protect\rho +q$ (=A) for $m=1$%
, $a=0.1$ and various values of $\protect\xi .$ The value of $\protect\xi $
indicates the deviation of the theory from standard gravity. From the figure
it is clear that by getting far from $R-$gravity the region in which WEC are
satisfied (i.e. $\protect\rho \geq 0$ and $\protect\rho +q\geq 0$) is
enlarged. We also note that having WEC satisfied, makes SEC satisfied as
well.}
\end{figure}
which can be satisfied. Fig. 1 displays the possible regions in which the
WEC are satisfied. Clearly by a fixed value for the Rindler acceleration
larger deviation from the standard $R$ gravity provides larger region of
satisfaction of WEC. Once the value of $\xi $ gets smaller and smaller the
region in which $\rho \geq 0$ and $\rho +q\geq 0$ gets narrower and
narrower. In the limit $\xi \rightarrow 0$ this region disappears completely.

In addition to the WEC one may also check the strong energy conditions (SEC)
i.e., (i) $\rho \geq 0$ (ii) $\rho +q\geq 0$ and (iii) $\rho +p+2q\geq 0.$
The first two conditions have already been considered in WEC and the third
condition becomes effectively equivalent with $q\geq 0$, i.e., 
\begin{equation}
-6a\xi r\ln \left( \frac{12a}{r}\right) -\left( \xi +2a\right) \leq 0,
\end{equation}%
which upon (20) is satisfied trivially.

In conclusion, the Grumiller metric \cite{4,5} would be physically
acceptable (from energy point of view) if instead of $R-$gravity we adopt $%
f\left( R\right) =R-12a\xi \ln \left\vert R\right\vert $ gravity. Herein $a$
is just the Rindler acceleration and $\xi $ is a parameter which shows the
deviation of the new $f(R)$ gravity from the standard gravity. The external
source consists of a fluid with an energy-momentum given by (2). The
pressure of the fluid is negative by choice so that it plays the role of
dark energy. The interesting feature of the Rindler modified Schwarzschild
geometry \cite{4,5} is that at large distances (i.e. outside the galaxy)
there exists still an effective mass to yield nearly flat rotation curves.
This result is irrelevant to whether the so-called Pioneer anomaly is a
genuine case or not. Clearly our model excludes the flat space (i.e. $R=0$)
but becomes applicable to spacetimes with $0<\left\vert R\right\vert <\infty
.$ Given the particular fluid source it is the unique $f(R)$ that yields the
Grumiller metric \cite{4,5} and satisfies WEC and SEC. It remains open,
however, to explore new forms of $f(R)$ with alternative energy-momenta
other than the one found here so that the same Rindler term will result with
energy conditions satisfied. This will be part of our future project.

\end{document}